\documentclass[11pt,a4paper]{amsart}

\usepackage{amsmath,amssymb,amsfonts,amsthm,amscd,yhmath,esint}
\usepackage[left=25mm,top=30mm,bottom=25mm,right=25mm]{geometry}
\usepackage[utf8]{inputenc}
\usepackage[english]{babel}
\usepackage[fixlanguage]{babelbib}
\usepackage{cmap}            
\usepackage[pdfdisplaydoctitle = true,
    colorlinks = true,
    urlcolor = blue,
    citecolor = blue,
    linkcolor = blue,
    pdfstartview = FitH,
    pdfpagemode = UseNone,
    bookmarksnumbered = true,
    unicode=true]{hyperref}  
\usepackage{hyperxmp}
\usepackage{bm}
\usepackage{calrsfs}
\usepackage[all]{xy}
\usepackage{float}
\usepackage{tensor}
\usepackage{booktabs}
\usepackage{todonotes}
\usepackage{cite}           
\usepackage{dutchcal}
\usepackage{enumitem}       
\usepackage{aliascnt}       
\usepackage{ifthen}         

\graphicspath{{figures/}}

\hypersetup{
    pdftitle={The Haar measure in solid mechanics},
    pdfauthor={Clément Ecker and Boris Kolev},
    pdfsubject={MSC 2020: 74B20; 74A05; 74A20; 58A10},
    pdfkeywords={Haar measure; Lie groups; Invariant theory; Mathematics of solid mechanics},
    pdflang=fr
}

\newcommand{\NewTheorem}[4][none]{

    \ifthenelse{\equal{#1}{none}}{ 
        \newaliascnt{#2}{#4}
        \newtheorem{#2}[#2]{#3}
        \aliascntresetthe{#2}
    }{ 
        \newtheorem{#2}{#3}[#1]
    }
    \expandafter\def\csname #2autorefname\endcsname{#3} 
}

\NewTheorem[section]{thm}{Theorem}{}

\NewTheorem{cor}{Corollary}{thm}
\NewTheorem{lem}{Lemma}{thm}
\NewTheorem{prop}{Proposition}{thm}

\theoremstyle{definition}
\NewTheorem{defn}{Definition}{thm}
\NewTheorem{exam}{Example}{thm}

\theoremstyle{remark}
\NewTheorem{rem}{Remark}{thm}

\numberwithin{equation}{section}

\newcommand{\NN}{\mathbb{N}}                
\newcommand{\RR}{\mathbb{R}}                
\newcommand{\CC}{\mathbb{C}}                

\newcommand{\HH}{\mathbb{H}}                            
\newcommand{\Sym}{\mathbb{S}}                           

\newcommand{\GL}{\mathrm{GL}}               
\newcommand{\OO}{\mathrm{O}}                
\newcommand{\SO}{\mathrm{SO}}               
\newcommand{\SU}{\mathrm{SU}}               

\newcommand{\g}{\mathfrak{g}}

\newcommand{\oo}{\mathfrak{o}}
\newcommand{\so}{\mathfrak{so}}
\newcommand{\su}{\mathfrak{su}}

\newcommand{\xx}{\mathbf{x}}

\newcommand{\ee}{\bm{e}}

\newcommand{\nn}{\bm{n}}

\newcommand{\abs}[1]{\left\vert#1\right\vert}
\newcommand{\set}[1]{\left\{#1\right\}}

\DeclareMathOperator{\tr}{Tr} %
\newcommand{\dd}{{\mathrm d}}

\DeclareMathOperator{\Rey}{R}
\DeclareMathOperator{\Tr}{Tr} %

\DeclareMathOperator{\im}{Im} %
\begin{document}

\title{The Haar measure in solid mechanics}

\author{C. Ecker}
\address[Clément Ecker]{Université Paris-Saclay, ENS Paris-Saclay, LMPS - Laboratoire de Mécanique Paris-Saclay, 91190, Gif-sur-Yvette, France}
\email{clement.ecker@ens-paris-saclay.fr}

\author{B. Kolev}
\address[Boris Kolev]{Université Paris-Saclay, ENS Paris-Saclay, CNRS,  LMT - Laboratoire de Mécanique et Technologie, 91190, Gif-sur-Yvette, France}
\email{boris.kolev@ens-paris-saclay.fr}

\date{\today}%


\begin{abstract}
  This article addresses the overlooked but crucial role of the Haar measure in solid mechanics, a concept well-established in mathematical literature but frequently misunderstood by mechanicians. The aim is to provide practical insights and methodologies for the application of the Haar measure in various mechanical scenarios. The article begins with an introduction to the Haar measure, underlying its significance and a theoretical foundation for its definition and formulas. Moving beyond mathematical abstraction, the core of the article relies on practical applications, including the computation of the Haar measure for the orthogonal transformations of the 2D and 3D space under several commonly used parametrisations, an application about invariant theory in mechanics and sampling on a group orbit. These applications are presented with practical examples, enabling mechanicians to integrate the Haar measure into their research seamlessly. The article caters to a broad audience, with sections designed for both introductory comprehension and in-depth exploration. Our goal is to equip mechanicians with a valuable tool, bridging the gap between mathematical theory and practical applications in the field of mechanics.
\end{abstract}


\maketitle

\tableofcontents

\section{Introduction}
\label{sec:introduction}

The \emph{Haar measure}, a standard notion in mathematics~\cite{Der2004,Stu2008,Fol1999,DCP2021}, is less familiar among mechanicians~\cite{dML+2024}, despite its critical role in problems involving rotation groups or orthogonal transformations in 2D and 3D spaces. The mechanical community have not extensively acknowledged this concept, as it is confirmed by its absence in several influential works~\cite{Bra2018,Pic2021,Cow1989,XAD2024}. One key challenge mechanicians face is the lack of accessible, practical methodologies for computing the \emph{Haar measure} for specific groups and parametrisations. A challenge compounded by mathematical literature that often omits explicit examples~\cite{Der2004,Stu2008,Fol1999} or do not always offer a straightforward path for non-experts~\cite{DCP2021, Ste1994}. Consequently, some mechanical texts about homogenisation in random media~\cite{Pic2021} or for the sampling of random rotations~\cite{Bra2018} fail to even mention the \emph{Haar measure}, despite being a prior knowledge in the referenced mathematical literature~\cite{Sch1997}. Besides, other fields like theoretical and quantum physicists are more familiar with this concept, where its use is well documented in introductory and advanced texts~\cite{Mel2023,Rum2002}.

The \emph{Haar measure} is the ``right'' integration measure on a Lie group, meaning it is invariant under the group’s operations, making it essential for integrating functions over the group. When working with groups of orthogonal transformations, such as rotations in 3D, the \emph{Haar measure} becomes indispensable, particularly in fields like homogenisation and rotation sampling~\cite{Ste1994,Rum2002}. Even if the \emph{Haar measure} is essentially unique, the real difficulty lies in obtaining a computable formula for the measure using local coordinate systems (or parametrisation) on the group, such as the Euler angles for 3D rotations.

Although critical in fields like invariant theory~\cite{Der2004,Stu2008,DCP2021,Tau2022,OKA2017,CDV2014}, where generating invariants under group actions requires a solid grasp of this concept, its use in the rest of the mechanical community has been limited. Yet, the \emph{Haar measure} is crucial in constructing invariant quantities in mechanics, with applications ranging from random fields and rotations to quantum mechanics, chemistry, and electromagnetism~\cite{DCP2021,Tau2022,Mel2023}.

The purpose of this article is to address this gap in the mechanical literature by providing a systematic, practical guide to computing and applying the \emph{Haar measure}. In particular, we will focus on deriving explicit formulas for the measure under commonly used parametrisations, such as the polar chart, Euler angles, and quaternions for the group of orthogonal transformations in 3D.

\subsection*{Outline}

This article is organised as follows. In \autoref{sec:Haar-measure}, we introduce the theoretical background on the \emph{Haar measure} and present the necessary formulas. We provide the main formula in \autoref{thm:FormulaHaar} which will be used as a systematic method for computing the \emph{Haar measure} under a given parametrisation of a given group. In \autoref{sec:SO2} and \autoref{sec:SO3}, we apply this methodology to the group of orthogonal transformations in 2D and 3D under the usual parametrisations: Euler angles, Polar angles and Quaternions. We will also link each formula to examples from mechanical applications. Finally, in \autoref{sec:Reynolds}, we illustrate the utility of the \emph{Haar measure} in invariant theory, including its use in computing the dimensions of isotropic and hemitropic tensors of any order in 3D (see \autoref{tab:dimiso3}), and in the study of a random variable over an orbit of a group action.

\subsection*{Notations}

The orthogonal group $\OO(D)$ is the subset of matrices $Q$ in $\mathrm{M}_{D}(\RR)$ such that $QQ^{\top} = Q^{\top}Q = I$ (matrix identity). The special orthogonal group $\SO(D)$ is the subgroup of $\OO(D)$ of matrices $Q$ with $\det Q = 1$. Both of these groups are closed subgroups of the general linear group $\GL(D)$ and are thus Lie groups. Their Lie algebras are
\begin{equation*}
  \oo(D) = \so(D) = \set{A\in\mathrm{M}_{D}(\RR);\; A^{\top} = -A}.
\end{equation*}
In 3D, there exists an accidental isomorphism $j$ between $\RR^3$ and $\so(3)$ given by
\begin{equation}
  \label{eq:isomorph3D}
  j:\begin{pmatrix}
    x_1 \\ x_2 \\ x_3
  \end{pmatrix} \in \RR^3 \to \begin{pmatrix}
    0 & x_3 & -x_2 \\ -x_3 & 0 & x_1 \\ x_2 & -x_1 & 0
  \end{pmatrix}\in\so(3),
\end{equation}
and such that\footnote{$\times$ is the usual cross product between vectors of $\RR^3$.}
\begin{equation*}
  [j(\xx_{1}),j(\xx_{2})] = j(\xx_{1} \times \xx_{2}),\quad \xx_1,\xx_2\in\RR^3.
\end{equation*}
A natural basis for $\so(3)$ is thus given by $(\xi_i:=j(\ee_i))_{i=1}^3$, where $\ee_i$ is the canonical basis of $\RR^{3}$. We will use the following notations.
\begin{enumerate}
  \item The rotation by angle $\alpha\in\RR$ around the axis generated by the unit vector $\nn$ is given, using the Olinde--Rodrigues formula
        \begin{equation}
          \label{eq:OlindeRodrigue}
          R(\nn, \alpha) := \exp (\alpha j(\nn)) = I + (\sin \alpha) j(\nn) + (1 - \cos \alpha)j(\nn)^{2}.
        \end{equation}
  \item The reflection through the plane normal to the unit vector $\nn$ is given by
        \begin{equation*}
          \pi(\nn)\xx := \xx - 2\langle \xx, \nn \rangle \nn.
        \end{equation*}
\end{enumerate}

\section{Haar measure on compact Lie groups}
\label{sec:Haar-measure}

Let us start with a finite group $G$ endowed with a probability measure $\mu$. Then, the expected value $E[f]$ of a random variable $f:G\to\RR$ is written as
\begin{equation*}
  E[f] = \sum_{g\in G} f(g)\mu(g), \quad \text{where} \quad \sum_{g\in G} \mu(g) = 1.
\end{equation*}
The measure $\mu$ is \emph{left-invariant} if $\mu(hg) = \mu(g)$ for all $h,g \in G$, it is right-invariant if $\mu(gh) = \mu(g)$ for all $h,g \in G$. It is straightforward to show that $\mu$ is left-invariant (or right-invariant) if and only if
\begin{equation*}
  \mu(g) = \frac{1}{\abs{G}},
\end{equation*}
where $\abs{G}$ is the cardinal of the finite group $G$. Such a probability measure, uniquely defined, is moreover \emph{bi-invariant}, meaning that it is both left and right invariant.

However, extending this concept to a non-finite group is less straightforward. In its more general form, a \emph{Haar measure} is defined for a locally compact topological group $G$ as a left (or right) invariant measure on the Borel algebra of $G$, and it is known to be uniquely defined up to a scaling factor~\cite{Wei1940,Car1940}. The simplest example of a Haar measure is the \emph{Lebesgue measure} on the (non compact) Abelian group $(\RR^{D},+)$. The Lebesgue measure is obviously bi-invariant, because $(\RR^{D},+)$ is a commutative group. This might not be the case in general for a non commutative locally compact group $G$. However, on a compact group $G$, a Haar measure can be shown to be bi-invariant and finite~\cite{Wei1940,Car1940,Fol1999}. Hence normalising it by fixing its total mass to one, we conclude that there is a uniquely defined left (or right) probability measure $\mu_{G}$ on $G$ and that $\mu_{G}$ is bi-invariant.

For a compact Lie group $G$ of dimension $d\in\NN$, the construction of the Haar measure $\mu_{G}$ can be made moreover explicit using the concept of \emph{volume-form}, that is a nowhere vanishing $d$-form (a $d$-alternate covariant tensor field on $G$). A volume form can be integrated over open subsets and defines a Borel measure on $G$~\cite{KN1963}. The construction of a Haar measure is then simple: choose a non zero alternate $d$-form on $\g = T_{e}G$, the Lie algebra of $G$, and translate it to each tangent space $T_{g}G$, using either the left translation $L_{g}: h \mapsto gh$ or the right translation $R_{g}: h \mapsto hg$. This way, one obtains a bi-invariant volume form on $G$ and normalising it by fixing its total mass to one, one gets the Haar measure on $G$. The construction of the Haar probability measure on a compact Lie group $G$ reduces therefore to find a volume form $\mu_{G}$ on $G$ which satisfies
\begin{equation}\label{eq:charaHaar}
  L_g^* \, \mu_{G} = \mu_{G}, \quad (\forall g\in G) \quad \text{and} \quad  \int_{G} \mu_{G} = 1.
\end{equation}

The solution of this problem is theoretically easy. Indeed, let us introduce the (left) Maurer-Cartan form on $G$, which is a left-invariant differential one-form on $G$, with values in the Lie algebra $\g$ of $G$ and defined by~\cite{Car1940}
\begin{equation*}
  \theta(X_{g}) := T_{g}L_{g^{-1}}. X_{g}, \qquad X_{g} \in T_{g}G
\end{equation*}
Then, choosing a basis $(\xi_i)$ of $\g$ and denoting by $\theta^{i}$ (scalar one-forms), the components of $\theta$ in this basis, the solution of the problem is given by
\begin{equation*}
  \mu_{G} := C^{-1} \, \theta^{1} \wedge \theta^{2} \wedge \dotsb \wedge \theta^{d}, \quad \text{where} \quad C := \int_{G} \theta^{1} \wedge \theta^{2} \wedge \dotsb \wedge \theta^{d},
\end{equation*}
and $\wedge$ is the exterior product (or wedge product) on differential forms~\cite{KN1963}.

The real difficulty is to obtain a computable formula, using a local coordinate system or some parametrisation of the group $G$. In the sequel, we will restrict to the case where $G$ is a \emph{matrix group}, which will be the case in the majority of the applications. This means that $G$ is a compact subgroup of $\GL_D(\RR)$, the group of $D \times D$ invertible matrices. In that case, the Maurer-Cartan form is simply written as $\theta(X_{g}) := g^{-1}X_{g}$.

Given a local coordinate system $(u^{i})$ on $G$, we denote by $(\partial u^{i})$, the corresponding basis of the tangent space $T_{g}G$, and by $(\dd u^{i})$ its dual basis. This local chart $(u^{i})$ defines a local parametrisation of $G$ (which, in practice, covers usually a dense open set in $G$)
\begin{equation*}
  p\colon U\subset\RR^d\to G \subset \GL_D(\RR) \subset \mathrm{M}_{D}(\RR),
\end{equation*}
where $\mathrm{M}_{D}(\RR)$ is the vector space of square matrices of size $D$. Then, the local expression of the Haar measure $\mu_{G}$ is given by
\begin{equation*}
  (p^*\mu_{G})_u = k(u)\, \dd u^1\wedge\dots\wedge\dd u^d,
\end{equation*}
where
\begin{align*}
  k(u) & = (p^*\mu_{G})_u(\partial u^1,\dots,\partial u^d)                                                                                                                                                                                          \\
       & = (\mu_{G})_{p(u)}(T_up \cdot \partial u^1, \dots, T_up \cdot \partial u^d )                                                                                                                                                               \\
       & = (\mu_{G})_{p(u)}\left(\frac{\partial p}{\partial u^1}, \dotsc ,\frac{\partial p}{\partial u^d}\right)                                                                                                                                    \\
       & = C^{-1} \, (\theta^i\wedge\dots\wedge\theta^d)_{p(u)}\left(\frac{\partial p}{\partial u^1}, \dotsc ,\frac{\partial p}{\partial u^d}\right) = C^{-1} \, \det\left( \theta^i_{p(u)} \left( \frac{\partial p}{\partial u^j} \right) \right).
\end{align*}
We will summarise this result in the following theorem.

\begin{thm}\label{thm:FormulaHaar}
  Let $p:U\to G$ be a local coordinates system on $G$, $(\xi_i)$, be a basis of the Lie algebra $\g$ of $G$, and $(\theta^{i})$ be the components of the left-invariant Maurer-Cartan form on $G$. Then the local expression of the Haar measure $\mu_{G}$ on $G$ is given by
  \begin{equation*}
    (p^*\mu_{G})_u = k(u)\,\dd u^1\wedge\dots\wedge\dd u^d, \quad \text{with}\quad k(u)= C^{-1} \det\left( \theta_{p(u)}^i\left( \frac{\partial p}{\partial u_j} \right) \right),
  \end{equation*}
  where $C$ is a constant to be determined such that $\mu_{G}$ is a probability measure on $G$.
\end{thm}

\begin{rem}
  Introducing the Frobenius scalar product on $\mathrm{M}_{D}(\RR)$
  \begin{equation*}
    \langle A,B \rangle = \frac{1}{2}\Tr(A B^{T}) = \frac{1}{2} (A : B),
  \end{equation*}
  and assuming that the basis $(\xi_i)$ is orthonormal, the expression of $k$ in \autoref{thm:FormulaHaar} reduces to,
  \begin{equation}
    \label{eq:matrixGroup}
    k(u)= C^{-1} \det \left\langle p(u)^{-1} \frac{\partial p}{\partial u_j} , \xi_i \right\rangle .
  \end{equation}
\end{rem}

\begin{rem}
  Given two different parametrisations of $G$,
  \begin{equation*}
    p_{1} \colon U\subset\RR^d\to G, \quad \text{with} \quad ({p_{1}}^*\mu_{G})_u = k_1(u)\, \dd u^1\wedge\dots\wedge\dd u^{d},
  \end{equation*}
  and
  \begin{equation*}
    p_{2} \colon W\subset\RR^d\to G, \quad \text{with} \quad ({p_{2}}^*\mu_{G})_w = k_2(w)\, \dd w^1\wedge\dots\wedge\dd w^{d},
  \end{equation*}
  we introduce the change of charts $\varphi := {p_{2}}^{-1} \circ p_{1}$
  \begin{equation*}
    \xymatrix{
    U \ar[dd]_{\varphi}\ar@/^/[dr]^{p_{1}} \\
    & G \\
    W\ar@/_/[ur]_{p_{2}}
    }
  \end{equation*}
  and we have then
  \begin{equation*}\label{eq:FormulaChangeCoord}
    k_1(u) = J_\varphi(u)k_{2}(\varphi(u)),
  \end{equation*}
  where $J_\varphi(u)$ is the determinant of the Jacobian of $\varphi$ at $u$.
\end{rem}

\section{Haar measure for \texorpdfstring{$\SO(2)$}{SO(2)} and \texorpdfstring{$\OO(2)$}{O(2)}}
\label{sec:SO2}

A first application, will be to apply this result to the rotation group of the plane $\SO(2) \subset \mathrm{M}_{2}(\RR)$. This group can be parametrised by the angle of the rotation $\alpha\in[0;2\pi[$, leading to the chart
\begin{equation*}
  p\colon [0;2\pi[\to \SO(2), \qquad \alpha \mapsto
  \begin{pmatrix}
    \cos\alpha & -\sin\alpha \\
    \sin\alpha & \cos\alpha
  \end{pmatrix}
\end{equation*}
Its Lie algebra, $\so(2)$, is the vector space of skew symmetric $2\times 2$ matrices and is spanned by the unit vector
\begin{equation*}
  \xi_1 =
  \begin{pmatrix}
    0 & -1 \\
    1 & 0
  \end{pmatrix}.
\end{equation*}
Therefore, \autoref{eq:matrixGroup} becomes,
\begin{align*}
  k(\alpha) & = -\frac{C^{-1}}{2}\tr \left(p(-\alpha)\frac{\partial p}{\partial \alpha}\xi_1 \right) \\
            & = -\frac{C^{-1}}{2} \tr \left(
  \begin{pmatrix}
      \cos\alpha  & \sin\alpha \\
      -\sin\alpha & \cos\alpha
    \end{pmatrix}
  \begin{pmatrix}
      -\sin\alpha & -\cos\alpha \\
      \cos\alpha  & -\sin\alpha
    \end{pmatrix}
  \begin{pmatrix}
      0 & -1 \\
      1 & 0
    \end{pmatrix} \right)  = C^{-1},
\end{align*}
where $C$ is determined by the condition
\begin{equation*}
  \int_{0}^{2\pi} k(\alpha) \dd\alpha = \int_{0}^{2\pi} C^{-1} \dd\alpha = 1.
\end{equation*}
Therefore, we obtain the usual formula for the \emph{Haar measure} of $\SO(2)$ (for the map $p:\alpha\in[0;2\pi[\to p(\alpha)\in\SO(2)$) as
\begin{equation*}
  (p^*\mu_{\SO(2)})_\alpha = \frac{\dd\alpha}{2\pi}.
\end{equation*}

To illustrates \autoref{eq:FormulaChangeCoord}, let us consider a second parametrisation of $\SO(2)$ given by
\begin{equation*}
  \tilde{p}\colon [-\pi;\pi[\to \SO(2), \qquad \beta \mapsto
  \begin{pmatrix}
    \cos\beta & -\sin\beta \\
    \sin\beta & \cos\beta
  \end{pmatrix}.
\end{equation*}
the change of chart is then just
\begin{equation*}
  \varphi \colon [0;2\pi[ \to [-\pi;\pi[, \qquad \alpha \mapsto \beta = \alpha-\pi ,
\end{equation*}
and we have
\begin{equation*}
  (\tilde{p}^*\mu_{\SO(2)})_\beta = \tilde{k}(\beta)\dd\beta = J_{\varphi^{-1}}(\beta)k(\varphi^{-1}(\beta))\dd\beta = 1k(\beta+\pi)\dd\beta=\frac{\dd\beta}{2\pi}.
\end{equation*}

The full orthogonal group of the plane $\OO(2)$ is composed by the rotations and the axial reflections. It can be written as
\begin{equation*}
  \OO(2) = \SO(2) \sqcup \sigma\SO(2), \quad \text{where } \sigma :=
  \begin{pmatrix}
    -1 & 0 \\
    0  & 1
  \end{pmatrix}.
\end{equation*}
Therefore, integrating a dummy function $f:\OO(2)\to \RR$ on $\OO(2)$ can be expressed as,
\begin{equation}
  \label{eq:integO2}
  \int_{\OO(2)} f(h)\, \mu_{\OO(2)} = \frac{1}{2} \left[\int_{\SO(2)}f(g)\,\mu_{\SO(2)} + \int_{\SO(2)}f(\sigma g)\,\mu_{\SO(2)}\right].
\end{equation}

\section{Haar measure for \texorpdfstring{$\SO(3)$}{SO(3)} and \texorpdfstring{$\OO(3)$}{O(3)}}
\label{sec:SO3}

As if it has not already been underlined enough, the \emph{Haar measure} for a group $G$ is a unique object but can only be described, for computation purposes, through a chart $p:U\to G$ of the group. This chart is not unique and the expression of the measure $p^*\mu_{G}$ in this chart is of course depending on it. In solid mechanics, several parametrisation of the rotations group are commonly used and it can be difficult to find the proper \emph{Haar measure} for each parametrisation especially if the parametrisation is not fully explicit~\cite{Bra2018,GPS2008}. The goal in this section is to provide a method of computation and formulas for the \emph{Haar measure} corresponding to three commonly used charts: the polar chart, the Euler (or Cardan) angles and the quaternions.

\subsection{The polar chart}
\label{subsec:polar-chart}

\begin{figure}[h]
  \centering
  \includegraphics[width=0.3\linewidth]{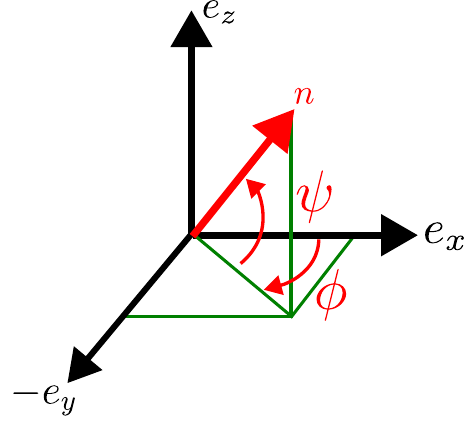}
  \caption{Spherical coordinates to represent a unit vector.}
  \label{fig:SphericalCoor}
\end{figure}

Let us choose first a common chart~\cite{Bra2018,DCP2021} in which a rotation $R(\nn,\alpha)$ is parametrised through the choice of an angle $\alpha\in[0,\pi[$ and a unit vector $\nn$ in the sphere $S^2$ and using a spherical coordinate system (see \autoref{fig:SphericalCoor}),
\begin{equation*}
  (\phi,\psi)\in [0,2\pi[\times\left[\frac{-\pi}{2},\frac{\pi}{2}\right]\mapsto \nn =
  \begin{pmatrix}
    \cos \psi \cos \phi \\
    \cos \psi \sin \phi \\
    \sin \psi
  \end{pmatrix}
\end{equation*}
and where
\begin{equation*}
  p:(\phi,\psi,\alpha)\in U = [0,2\pi[\times\left[\frac{-\pi}{2},\frac{\pi}{2}\right]\times[0,\pi]\mapsto R(\nn(\phi,\psi),\alpha)=R(\nn,\alpha)\in \mathrm{M}_3(\RR),
\end{equation*}
in which $R(\nn,\alpha)$ is the matrix expression deduced by the Olinde-Rodrigues formula (see \autoref{eq:OlindeRodrigue}),
\begin{equation*}
  R(\nn, \alpha) := \exp (\alpha j(\nn)) = I + (\sin \alpha) j(\nn) + (1 - \cos \alpha)j(\nn)^{2},
\end{equation*}
with $j$, the isomorphism defined in \autoref{eq:isomorph3D}. The \emph{Haar measure} $p^*\mu_{\SO(3)}$ of $\SO(3)$ in this parametrisation is given by
\begin{equation*}
  u=(\phi,\psi,\alpha), \quad (p^*\mu_{\SO(3)})_u = k(\phi,\psi,\alpha)\,\dd\phi\,\dd\psi\,\dd\alpha.
\end{equation*}
We use \autoref{eq:matrixGroup} (derived from \autoref{thm:FormulaHaar}) with $\xi_i=j(\ee_i)$, and we get
\begin{equation*}
  k(\phi,\psi,\alpha) = \det \left\langle \tau_i(\phi,\psi,\alpha),j(\ee_j) \right\rangle, \quad\text{with}\quad \tau_i = R(\nn(\phi,\psi),-\alpha)\frac{\partial}{\partial u^i} \left( R(\nn(\phi,\psi),\alpha) \right)\in\so(3).
\end{equation*}
Let $T$ be the components' matrix of the vectors $j^{-1}(\tau_i)\in\RR^3$ in the canonical basis of $\RR^3$, we have thus
\begin{equation*}
  k(\phi,\psi,\alpha) = \det T.
\end{equation*}

\begin{rem}
  Computing the matrix $T$ and taking its determinant is a tedious computation by hand but straightforward using a symbolic computation software. However, for a careful reader one can notice that $\tau_3=R(\nn(\phi,\psi),-\alpha)\frac{\partial}{\partial \alpha} \left( R(\nn(\phi,\psi),\alpha) \right)=j(\nn(\phi,\psi))$ and $\frac{\partial \nn}{\partial \psi}\cdot \nn=0$. Therefore one can try to compute the matrix $T$ in the orthonormal basis $(\nn,\frac{\partial \nn}{\partial \psi},n\wedge \frac{\partial \nn}{\partial \psi})$ which is doable by hand.
\end{rem}
We conclude that,
\begin{equation}
  \label{eq:HaarSO3Polar}
  u=(\phi,\psi,\alpha)\in U,\quad (p^*\mu_{\SO(3)})_u=\frac{1}{2\pi^2}\cos\psi\cos^2\left(\frac{\alpha}{2}\right)\,\dd\phi\,\dd\psi\,\dd\alpha.
\end{equation}

\begin{rem}
  \label{rem:recoverParam}
  As it is custody when working with rotation parametrisations it can be difficult to identify in the literature which parametrisation has been used. For example in~\cite{DCP2021}, they are using a similar parametrisation but without detailing the construction of the vector $\nn$ from spherical coordinates neither specifying the ranges of the three angles. However, from \autoref{eq:HaarSO3Polar} it is possible to identify the ranges of angles by applying simple translations (using \autoref{eq:FormulaChangeCoord}) to the angles.
\end{rem}

\subsection{Euler angles}
\label{title:SO3HaarEuler}

\begin{figure}[h]
  \centering
  \includegraphics[width=0.9\linewidth]{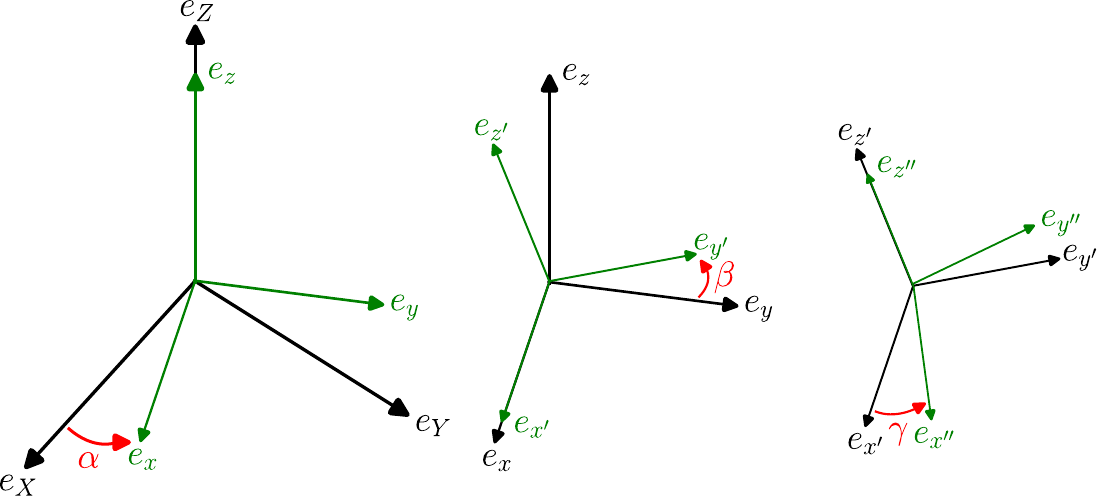}
  \caption{Representation of a rotation using Euler angles.}
  \label{fig:EulerAngleFig}
\end{figure}

Another famous parametrisation of $\SO(3)$ uses the Euler angles~\cite{Bra2018,GPS2008,dML+2024}. Given an initial orthonormal basis $\{\ee_X, \ee_Y,\ee_Z\}$ on which we apply (see \autoref{fig:EulerAngleFig}):
\begin{enumerate}
  \item a first rotation of angle $\alpha\in[-\pi;\pi]$ around the $\ee_Z$ axis leading to a new basis $\{\ee_{x}, \ee_{y}$, $\ee_{z} = \ee_{Z}\}$;
  \item a second rotation of angle $\beta\in[0;\pi]$ around the $\ee_{x}$ axis leading to a new basis $\{\ee_{x'} = \ee_{x}$, $\ee_{y'}, \ee_{z'}\}$;
  \item a third rotation of angle $\gamma\in[-\pi;\pi]$ around the $\ee_{z'}$ axis leading to a new basis $\{\ee_{x''}, \ee_{y''}$, $\ee_{z''} = \ee_{z'}\}$.
\end{enumerate}
This corresponds to the following chart
\begin{gather*}
  p:u=(\alpha,\beta,\gamma)\in U=[-\pi;\pi]\times[0;\pi]\times[-\pi;\pi] \to R(u), \\
  R(u) =
  \begin{pmatrix}
    \cos\alpha\cos\gamma-\cos\beta\sin\alpha\sin\gamma & -\cos\alpha\sin\gamma-\cos\beta\cos\gamma\sin\alpha & \sin\alpha\sin\beta  \\
    \cos\gamma\sin\alpha+\cos\alpha\cos\beta\sin\gamma & \cos\alpha\cos\beta\cos\gamma-\sin\alpha\sin\gamma  & -\cos\alpha\sin\beta \\
    \sin\beta\sin\gamma                                & \cos\gamma\sin\beta                                 & \cos\beta
  \end{pmatrix}.
\end{gather*}

The exact same methodology can be applied in order to compute the \emph{Haar measure} $p^*\mu_{\SO(3)}$ (using \autoref{eq:matrixGroup}) in this chart,
\begin{equation*}
  u=(\alpha,\beta,\gamma), \quad (p^*\mu_{\SO(3)})_u = k(\alpha,\beta,\gamma)\dd \alpha\dd\beta\dd\gamma.
\end{equation*}
We express \autoref{eq:matrixGroup} (derived from \autoref{thm:FormulaHaar}) in the case where $\xi_i=j(\ee_i)$. However, this time we do not have an easy way to compute $R(u)^{-1}$ (unlike with the polar coordinates where $R(\nn(\phi,\psi),\alpha)^{-1}=R(\nn(\phi,\psi),-\alpha)$), and to build the components' matrix $T$ of the three vectors $j^{-1}(\tau_i)\in\RR^3$. We have anyway
\begin{equation*}
  k(\alpha,\beta,\gamma) = \det T, \quad\text{where}\quad \tau_i = R(\alpha,\beta,\gamma)^{-1}\frac{\partial R}{\partial u^i} (\alpha,\beta,\gamma)\in\so(3).
\end{equation*}
Using a symbolic computation software, we obtain the \emph{Haar measure} in this chart
\begin{equation}
  \label{eq:HaarEuler}
  u=(\alpha,\beta,\gamma)\in[-\pi;\pi]\times[0;\pi]\times[-\pi;\pi],\quad (p^*\mu_{\SO(3)})_u = \frac{1}{8\pi^2}\sin\beta\, \dd\alpha\,\dd\beta\,\dd\gamma.
\end{equation}

\begin{rem}
  As it is emphasised in the literature~\cite{Bra2018,GPS2008}, there are often ambiguities when using Euler angles. Indeed, there exist several conventions for the range of angles $\alpha,\beta,\gamma$, as well as for the axis of the rotations and it is not always clearly stated which convention is used. For example, the Cardan angles correspond to another choice of ranges and rotations combinations (see~\cite{GPS2008} for an extensive discussion on that topic). For instance, in~\cite{dML+2024}, the convention used is not explicitly mentioned but like in \autoref{rem:recoverParam}, it is possible by translating our angles, to recover their formula for the \emph{Haar measure}.
\end{rem}

\begin{rem}
  It can be noticed that we could have used \autoref{eq:FormulaChangeCoord} and the previous formula for the \emph{Haar measure} in polar coordinates in order to obtain this expression. However, it is usually easier (in some sense) to stick to the general \autoref{thm:FormulaHaar} because finding the change of coordinates for rotations often involves the definition of generalised (covering the whole domain) inverse of trigonometric functions that can be difficult to handle especially when differentiating (see~\cite{dML+2024}). It is why \autoref{eq:FormulaChangeCoord} is often more useful to recover ``relatively simple'' changes of variables like a change of range in the parameter rather than a completely different parametrisation.
\end{rem}

\subsection{Quaternions}

Another common practice when dealing with rotations is to use quaternions~\cite{Bra2018,Ste1994,GPS2008} to parameterize the rotation group $\SO(3)$. This parametrerization uses the group of unitary quaternion (quaternions of norm $1$). A unitary quaternion $q=w+xi+ yj+zk\in Q$ can be identified to a rotation $R(\nn,\alpha)$ through the following process:
\begin{enumerate}
  \item The rotation axis $\nn$ is given by $\nn=(x,y,z)$;
  \item The angle of rotation $\alpha$ is given by $\alpha=2\arccos w=2\arcsin\sqrt{x^2+y^2+z^2}$;
  \item The rotated vector $v^R$ of a vector $v\in\RR^3$ is given by the quaternion multiplications $qv\bar{q}$, in which :
        \begin{enumerate}
          \item A vector $v\in\RR^3$ is associated to a pure imaginary quaternion $v_xi+v_yj+v_zk$;
          \item $\bar{q}$ is the conjugate of $q$ given by $\bar{q}=w-xi-yj-zk$.
        \end{enumerate}
\end{enumerate}

Writing a quaternion of norm $1$ in terms of $w,x,y,z$ identifies the group of unitary quaternions with the 3-sphere $S^3$ in $\RR^{4} \simeq \CC^{2}$ and consequently the 3-sphere can be mapped to the group of rotations. However, this transformation is not injective, only surjective, in a sense that if $q\in Q$ identifies to $R(\nn,\alpha)$ then $-q\in Q$ identifies to $R(-\nn,-\alpha)=R(\nn,\alpha)$.

\begin{rem}
  In more rigorous terms, the three sphere $S^3 \subset Q$ has the structure of a Lie group noted $\SU(2)$ and the above construction must be analysed using the adjoint action of $\SU(2)$ on its Lie algebra $\su(2)$, but it is outside the scope of this article.
\end{rem}

Identifying a unit quaternion as a point in a 3-sphere already implies, if some computations wants to be performed, that a suitable parametrisation (and \emph{Haar measure}) has already been chosen to map the 3-sphere.

The most common way to do it is to use the hyperpolar parametrisation given by,
\begin{equation*}
  p:(\theta,\psi,\phi)\in U \to q=w+xi+ yj+zk\in Q\subset S^3 \subset \HH,
\end{equation*}
with
\begin{equation*}
  (\theta,\psi,\phi)\in U=[0;\pi]\times[0;\pi]\times[0;2\pi],\quad \left\{
  \begin{array}{ll}
    w & =\cos\theta                 \\
    x & =\sin\theta\cos\psi         \\
    y & =\sin\theta\sin\psi\cos\phi \\
    z & =\sin\theta\sin\psi\sin\phi \\
  \end{array} \right.
\end{equation*}

Since $\SU(2)$ (the set of unitary quaternions) is a Lie group and its Lie algebra $\su(2)$ is the space of pure imaginary quaternions, we can take the basis $(\xi_1,\xi_2,\xi_3)=(i,j,k)$. Similarly to \autoref{subsec:polar-chart} and \autoref{title:SO3HaarEuler}, we directly apply \autoref{thm:FormulaHaar} (\autoref{eq:matrixGroup} also applies in that case) but the computations can be mainly performed by hand. The derivatives of the map $p$ are straightforward,
\begin{equation*}
  \left\{ \begin{array}{ll}
    \frac{\partial p}{\partial \theta} & = -\sin\theta+i\cos\theta\cos\psi+j\cos\theta\sin\psi\cos\phi+k\cos\theta\sin\psi\sin\phi \\
    \frac{\partial p}{\partial \psi}   & = -i\sin\theta\sin\psi+j\sin\theta\cos\psi\cos\phi+k\sin\theta\cos\psi\sin\phi            \\
    \frac{\partial p}{\partial \phi}   & = -j\sin\theta\sin\psi\sin\phi+k\sin\theta\sin\psi\cos\phi.
  \end{array} \right.
\end{equation*}
The elements $\tau_i:=p(\theta,\psi,\phi)^{-1}\frac{\partial p}{\partial u^i}$ of $\mathfrak{q}$ are computed using the quaternion multiplication and inverse (for a unitary quaternion $q\in Q$, $q^{-1}=\bar{q}$)
\begin{equation*}
  \left\{ \begin{array}{ll}
    \tau_\theta & = i\cos\psi+j\sin\psi\cos\phi+k\sin\psi\sin\phi                                                                                                          \\
    \tau_\psi   & = \sin\theta\bigl( -i\cos\theta\sin\psi+j(\cos\theta\cos\psi\cos\phi+\sin\theta\sin\phi)+k(\cos\theta\cos\psi\sin\phi-\sin\theta\cos\phi) \bigr)         \\
    \tau_\phi   & = \sin\theta\sin\psi\bigl( -i\sin\theta\sin\psi+j(\sin\theta\cos\psi\cos\phi-\cos\theta\sin\phi)+k(\cos\theta\cos\phi+\sin\theta\cos\psi\sin\phi) \bigr)
  \end{array} \right.
\end{equation*}
In the basis $(i,j,k)$ of $\su(2)$ the matrix $T$ expresses as
\begin{equation*}
  T =
  \begin{pmatrix}
    \cos\psi                      & \sin\psi\cos\phi                                                  & \sin\psi\sin\phi                                                  \\
    -\cos\theta\sin\theta\sin\psi & \sin\theta(\cos\theta\cos\psi\cos\phi+\sin\theta\sin\phi)         & \sin\theta(\cos\theta\cos\psi\sin\phi-\sin\theta\cos\phi)         \\
    -\sin^2\theta\sin^2\psi       & \sin\theta\sin\psi(\sin\theta\cos\psi\cos\phi-\cos\theta\sin\phi) & \sin\theta\sin\psi(\cos\theta\cos\phi+\sin\theta\cos\psi\sin\phi)
  \end{pmatrix}
\end{equation*}
Finally, using a symbolic computation software to evaluate the determinant of $T$ (and finding the normalisation factor $C$), the expression of the \emph{Haar measure} in this coordinate system is,
\begin{equation}
  u=(\theta,\psi,\phi)\in U=[0;\pi]\times[0;\pi]\times[0;2\pi],\quad (p^*\mu_{\SO(3)})_u = \frac{1}{2\pi^2}\sin^2\theta\sin\psi\dd\theta\dd\psi\dd\phi.
\end{equation}

\begin{rem}
  This expression is the same as the one which can be found in~\cite{Ste1994}.
\end{rem}

\subsection{Haar measure for \texorpdfstring{$\OO(3)$}{OO(3)}}

Extending the Haar measure to the full orthogonal group $\OO(3)$ is straightforward since the reasoning is exactly the same as in \autoref{eq:integO2} for $\OO(2)$,
\begin{equation*}
  \OO(3) = \SO(3)\sqcup(-I)\SO(3).
\end{equation*}
Therefore, integrating a dummy function $f:\OO(3)\to \RR$ on $\OO(3)$ can be expressed as,
\begin{equation}
  \label{eq:integO3}
  \int_{\OO(3)} f(h)\, \mu_{\OO(3)} = \frac{1}{2}\left[\int_{\SO(3)}f(g)\,\mu_{\SO(3)} + \int_{\SO(3)}f( -g)\,\mu_{\SO(3)}\right],
\end{equation}
and this formula can also be found in~\cite{DCP2021}.

\subsection*{Uses in solid mechanics}

Euler angles remain the most commonly used parametrisation when dealing with theoretical or analytical results. For instance, in~\cite{XAD2024}, an article on the homogenisation of porous metals (Gurson model) with randomly oriented spherical voids, and in~\cite{GZV2024}, an article about the homogenisation of concrete, both utilize the formula provided in \autoref{eq:HaarEuler}. However, both articles cite~\cite{And2004}, an article about molecular dynamics, without referencing or demonstrating the formula itself. A similar issue arises in~\cite{SKC2023}, where the formula is applied to the homogenisation of wave properties in random solid media but is not cited or discussed in depth. An exception can be found in~\cite{Fer2020}, where the polar parametrisation, described in \autoref{eq:HaarSO3Polar}, is used and justified. For numerical simulations, quaternions are more frequently employed~\cite{Bra2018,Sch1997}.

\section{Invariant theory: the Reynolds projector}
\label{sec:Reynolds}

After expressing the \emph{Haar measure} for different groups and different parametrisations and in order to emphasis its usability, we will illustrates its use in different scenarios.

Let $T$ be a tensor of interest describing a material property of order $n$ over $\RR^D$, therefore\footnote{For $T_{i_1\dots i_n}$ a tensor of order $n$ and $g_{ij}$ a change of frame, we have, $(g\star T)_{i_1\dots i_n}=g_{i_1j_1}\dots g_{i_nj_n} T_{j_1\dots j_n}$.} $g\star T$ represents its expression in a rotated basis and we say that $G$ acts on $T$. Now, $g\star T=T$ means that the tensor $T$ is invariant regarding the action of $g$ and that $g$ is a material symmetry of $T$ (the symmetry group $G$ of $T$ contains all the material symmetries) \cite{ZB1994}. In a context of homogenisation, one could be more interested in describing the (uniform or equiprobable) combination of those orientations leading to an equivalent tensor $\tilde{T}$,
\begin{equation}\label{eq:ReyDiscrete}
  \tilde{T} = \sum_{g\in G} \mu_{G}(g) \, g\star T ,
\end{equation}
in which $\mu_{G}(g)$ is the weight given to the rotation $g$. In this situation it is obvious that $\mu_{G}(g)=1/|G|$ in order to have an averaged tensor and more generally, it is natural to use the \emph{Haar measure} associated to the group since we want to attribute a uniform weight to all the elements of $G$. Consequently, the expression of \autoref{eq:ReyDiscrete} can be easily generalised to any compact group $G$ by,
\begin{equation}\label{eq:ReyConti}
  \tilde{T} = \int_{G} g\star T \, \mu_{G}, \quad \text{with $\mu_{G}$ the \emph{Haar} measure on the group $G$}.
\end{equation}

Since this expression can be adapted to any tensor and more generally to any (finite dimensional) vector space $V$ on which $G$ acts linearly, it is common to define an application, the \emph{Reynolds projector}~\cite{Der2004,Stu2008,Tau2022} to represent this averaging process. This application noted $R^{V}_{G}$ is built using the \emph{Haar measure} $\mu_{G}$ on $G$ and a given linear representation $\rho_V:G\to\GL(V)$ of $G$ on $V$. The \emph{Reynolds projector} is then the linear operator defined on $V$ by
\begin{equation}\label{eq:ReyDef}
  \Rey_{G}^{V}(v) = \int_{G} \rho_V(g) v \, \mu_{G}.
\end{equation}
This linear operator is indeed a \emph{linear projector}, meaning that $(\Rey_{G}^{V})^{2} = \Rey_{G}^{V}$.

The intuition initially developed in the introductory problem can be enlarged to a very broad concern in mechanics. Given a quantity of interest, commonly tensorial~\cite{TN1965}, and a group of transformations, commonly orthogonal, what can we tell about the invariance, or covariance, of those quantities under those transformations ? A vast and old literature~\cite{Wey1946,Riv1955} exists on those topics and makes a proficient use of mathematical theories developed in the 19th century. The humble objective of this section is to introduce the reader to a common tool which is defined using the \emph{Haar measure} making it a direct application of the previous section. Getting back to our \emph{Reynolds projector} of \autoref{eq:ReyDef}, it becomes a useful tool to study the set $V^{G}$ of invariant elements of $V$ under the action of $G$,
\begin{equation}\label{eq:invariantSet}
  V^{G} := \set{v\in V ;\; \rho_V(g)v=v, \, \forall g\in G}.
\end{equation}

Indeed, it is easy to verify that the \emph{Reynolds projector} is invariant under the action of $G$ ($\rho_V(g)\Rey^{V}_{G}(v)=R^{V}_{G}(v)$) leading that the image of the \emph{Reynolds projector} is exactly the set $V^{G}$ of invariant elements of $V$~\cite{Der2004}, confirming the interest of studying the application. In our illustration we will focus on determining the dimension of $V^{G}$ which is of great interest for mechanicians since it corresponds to the number of coefficients necessary to describe a tensor of interest when $G$ is the group of material symmetries~\cite{dML+2024,Tau2022,OKA2017}. For that, we will use the fact that $\Rey_{G}^{V}$ is a projector and use the \emph{trace formula}
\begin{equation}
  \label{eq:traceFormula}
  \dim V^{G} = \dim \im(\Rey_{G}^{V}) = \int_{G}\Tr(\rho_V(g))\,\mu_{G}.
\end{equation}

Usually, in mechanics, $V$ is a subspace of the tensor space $\otimes^{n}\RR^{D}$, defined by certain index symmetries~\cite{OKA2017} (the strain tensor or the elasticity tensor, for instance) and $G$ is a subgroup of $\GL_D(\RR)$. In that case $\rho_V$ can be taken as the restriction to $V$ of the tensorial representation $\rho_n:=\otimes^{n}\rho$ (denoted as $\rho_V={\rho_n}_{\vert V}$), where $\rho$ is the natural representation of $G$ on $\RR^{D}$. If no index symmetries are considered and $V=\otimes^{n}\RR^{D}$, we have $\Tr(\rho_V(g)) = \Tr(\rho(g))^{n}$ and we get
\begin{equation}
  \label{eq:ReyPlongement}
  \dim(V^{G}) = \int_{G} \Tr(\rho(g))^{n} \,\mu_{G}.
\end{equation}

\begin{rem}
  One should however be careful when $V$ is a proper subspace of $\otimes^{n}\RR^{D}$, defined by some index symmetries and which is obviously stable by $\rho_n$, meaning that
  \begin{equation*}
    \rho_n(g)w\in W, \qquad \forall g\in G, \quad \forall w\in W.
  \end{equation*}
  For example, if $G=\OO(D)$, and $V=\Sym_2(\RR^D)$, the space of symmetric second order tensors, which is stable by $\rho_{2}$ (where $\rho_{2}(g)v := g^{-1}vg$) since $(g^{-1}vg)^T=g^{-1}vg$. In that case, we have $\Tr {\rho_2}_{\vert \Sym_2(\RR^D)} \neq \Tr(\rho)^{2}$ which can be verified by analysing the dimensions of the operators.
\end{rem}

\subsection{First application: Isotropic and hemitropic tensors}

In solid mechanics, tensors are often invariant under the action of a group and the group of (special) orthogonal transformations are of particular interest. There is a debate for the definition of an isotropic tensor in 3D, if they are the tensors invariant under $\SO(3)$ or $\OO(3)$. We adopt the definition of~\cite{ZB1994} and hemitropic tensors are the ones invariant under $\SO(3)$ and isotropic ones are invariant under $\OO(3)$ (for example~\cite{AT1977} calls the tensors invariant under $\SO(3)$ isotropic).

Let $G=\SO(3)$ and $V=\otimes^{n}\RR^{3}$. To perform the calculations, we will use the polar chart $p$ introduced in \autoref{subsec:polar-chart} and the matrix notation on $\GL_3(\RR)$. In this situation \autoref{eq:ReyPlongement} will give the dimension of hemitropic (isotropic if $G=\OO(3)$) tensors of order $n$. With this chart, the matrix representation of a rotation $R(\nn,\alpha)$ is simple in the basis $\mathcal{B} := (\nn, \nn^\perp,\nn\wedge\nn^\perp)$ in which $\nn^\perp$ is a unitary vector normal to $\nn$. We get
\begin{equation*}
  [R(\nn,\alpha)]_{\mathcal{B}} =
  \begin{pmatrix}
    1 & 0          & 0           \\
    0 & \cos\alpha & -\sin\alpha \\
    0 & \sin\alpha & \cos\alpha
  \end{pmatrix}
\end{equation*}
leading to
\begin{equation*}
  \Tr(R(\nn,\alpha)) = 1+2\cos\alpha.
\end{equation*}
Therefore, a direct application of \autoref{eq:ReyPlongement} using the expression of the \emph{Haar measure} given by \autoref{eq:HaarSO3Polar}, leads to
\begin{equation*}
  D_n := \dim((\otimes^{n}\RR^3)^{\SO(3)}) = \frac{1}{2\pi^2}\int_{U}(1+2\cos\alpha)^{n}\cos\psi\sin^2\left(\frac{\alpha}{2}\right)\,\dd\phi\,\dd\psi\,\dd\alpha.
\end{equation*}
To compute this integral, we need to distinguish according to the parity of $n$, and we get
\begin{align*}
  D_{2m}   & = \sum_{k=0}^m\binom{2k}{k}\left[\frac{3}{2}\binom{2m}{2k}-\frac{1}{2}\binom{2m+1}{2k}\right]  ,                                \\
  D_{2m+1} & = \sum_{k=0}^m\binom{2k}{k}\left[\frac{3}{2}\binom{2m+1}{2k}-\frac{1}{2}\binom{2m+2}{2k}\right] - \frac{1}{2}\binom{2m+2}{m+1}.
\end{align*}
A very similar computation can be done for $G=\OO(3)$ by applying the integration over this group given by \autoref{eq:integO3}. Using the fact that $\Tr((-I)g) = -\Tr(g)$, we get
\begin{align*}
  \dim((\otimes^{n}\RR^3)^{\OO(3)}) & = \frac{1}{4\pi^2}\int_{u\in U}(1+(-1)^{n})(1+2\cos\alpha)^{n}\cos\psi\sin^2\left(\frac{\alpha}{2}\right)\dd\phi\dd\psi\dd\alpha \\
                                    & = \left\{ \begin{array}{ll}
                                                  D_{2m} & \text{if } n=2m,   \\
                                                  0      & \text{if } n=2m+1.
                                                \end{array} \right.
\end{align*}

\begin{table}[h]
  \centering
  \begin{tabular}{lccccccccccc}
    \toprule
    $n$                                 & 0 & 1 & 2 & 3 & 4 & 5 & 6  & 8  & 15    & 16      & 20         \\
    \midrule
    $\dim (\otimes^{n}\RR^3)^{\SO(3)} $ & 1 & 0 & 1 & 1 & 3 & 6 & 15 & 91 & 83097 & 227 475 & 13 393 689 \\
    $\dim (\otimes^{n}\RR^3)^{\OO(3)} $ & 1 & 0 & 1 & 0 & 3 & 0 & 15 & 91 & 0     & 227 475 & 13 393 689 \\
    \bottomrule
  \end{tabular}
  \caption{Dimensions of hemitropic and isotropic tensors.}
  \label{tab:dimiso3}
\end{table}

\begin{rem}
  A similar problem arises in material constitutive modelling : given several tensors of interest, is it possible to build a function, typically polynomial, depending on these tensors, which will be invariant under the group action ? For example in classical continuum mechanics, the point-wise energy of a system depends on several tensorial quantities (strains, electromagnetic field, temperature, etc) and one tries to find an energy function depending on those quantities that will be invariant under any frame changes. It is often not very difficult to build such a function however it is much more difficult to find a family of functions, preferably parameterised by a finite set of coefficients, such as any invariant function (of a certain type) can be written in that form. It exists a tool in invariant theory, the \emph{Hilbert series} of a graded commutative algebra, that allows to determine the number of independent coefficients necessary to describe such a family of functions. Indeed, the set of invariant polynomials under the action of a group forms an algebra and this series counts the dimension (the number of independent coefficients) of invariant polynomials of each degree. In the case of a Lie group the \emph{Hilbert series} can be computed a priori using the \emph{Molien-Weyl formula}~\cite[Sect 2.2]{Stu2008} which makes use of the \emph{Reynolds projector} but this topic is outside of the scope of the article.
\end{rem}

\subsection*{Uses in solid mechanics} The dimension of hemitropic tensors of order $n$ (\autoref{tab:dimiso3}), is also found in~\cite{AT1977}, which uses the same method but without detailed explanations, relying on the Euler angles parametrisation. Techniques based on the \emph{Molien-Weyl formula} have been particularly useful in~\cite{Tau2022} for generating energy densities in the study of magnetostriction.

\subsection{Second application: Uniform sampling of an orbit}

Since the \emph{Haar measure} is a uniform measure on a compact group $G$, it can be interested to build probability measures on orbits of a linear representation $\rho_V:G\mapsto\GL(V)$ of $G$ on a vector space $V$. Let $v\in V$ and let us consider
\begin{equation*}
  \label{eq:defVAR}
  X:g\in G\to \rho_V(g)v\in V,
\end{equation*}
as a (vector valued) random variable on $G$.

A natural question arises in this situation : if we sample $G$ uniformly (following the \emph{Haar measure}), what is the probability distribution of $X$ ? Let us study the $r$-moments $m_r(X)$ of $X$ that can be seen as a random vector on the probability space $(G,\sigma(G),\mu_{G})$,
\begin{align*}
  m_r(X) = E[\underbrace{X\otimes\dots\otimes X}_{r\text{ times}}] & = \int_{G} \Bigl(\otimes^r (\rho_V(g)v)\Bigr) \, \mu_{G}                      \\
                                                                   & = \int_{G} \Bigl(\otimes^r \rho_V(g)\Bigr)\left(\otimes^r v \right) \,\mu_{G} \\
                                                                   & = \Rey_{G}^{\otimes^r V}(\otimes^rv),
\end{align*}
in which $\sigma(G)$ is the Borel algebra of $G$ and $\Rey_{G}^{\otimes^rV}$ is the \emph{Reynolds projector} of $G$ on $\otimes^rV$ (see \autoref{eq:ReyDef} for the definition of the Reynolds projector). Noticeably, the appearance of the \emph{Reynolds projector} is not accidental and illustrates that sampling the group according to the \emph{Haar measure} (uniformly) creates a probability distribution of $X$ which only depends on invariant quantities (regarding the action of $G$ on $\otimes^rV$). Therefore, the two first useful quantities, the expectation and the covariance matrix are given by
\begin{align*}
  E[X]=m_1(X)                               & = \Rey_{G}^{V}(v),                                                         \\
  \text{Cov}(X)=m_1(X)\otimes m_1(X)-m_2(X) & = \Rey_{G}^{V}(v)\otimes\Rey_{G}^{V}(v)-\Rey_{G}^{V\otimes V}(v\otimes v).
\end{align*}

\begin{exam}
  In the case were $G=\SO(3)$ and $V=\Sym_2(\RR^3)$ the space of symmetric $3\times 3$ matrices, we have for $v_0\in V$ a diagonal matrix,
  \begin{equation*}
    v_0= \begin{pmatrix} \lambda_1 & 0 & 0 \\ 0 & \lambda_2 & 0 \\ 0 & 0 & \lambda_3 \end{pmatrix}, \quad E[X]= m_1(X) = \frac{\Tr(v_0)}{3} I = \frac{\lambda_1+\lambda_2+\lambda_3}{3} I,
  \end{equation*}
  because the dimension of isotropic matrices is one (see \autoref{tab:dimiso3}) and the trace operation is invariant regarding the application of the group. For the covariance we use a similar strategy, it is easy to show that\footnote{In which $\delta_{ij}$ and $\delta_{ijkl}$ are the Kronecker symbols : $\delta_{ij}=\left\{ \begin{array}{ll} 1&\text{if }i=j \\ 0&\text{else} \end{array}\right.$, $\delta_{ijkl}=\left\{ \begin{array}{ll} 1&\text{if }i=j=k=l \\ 0&\text{else} \end{array}\right.$}
  \begin{equation*}
    (\mathbb{J}_1, \mathbb{J}_2= I\otimes I) = (\delta_{ijkl}, \delta_{ij}\delta_{kl})
  \end{equation*}
  is a generating set of $(\Sym_2(\RR^3)\otimes \Sym_2(\RR^3))^{\SO(3)}$ and since the scalar product for fourth order tensors is invariant under the action of $\SO(3)$, we can express $m_2(X)$ as,
  \begin{equation*}
    m_2(X) = \frac{3\Tr(v_0^2)-\Tr(v_0)^2}{6}\mathbb{J}_1 + \frac{\Tr(v_0)^2-\Tr(v_0^2)}{6}\mathbb{J}_2,
  \end{equation*}
  thus,
  \begin{equation*}
    \text{Cov}(X)=\frac{3\Tr(v_0^2)-\Tr(v_0)^2}{6}\left(\frac{\mathbb{J}_2}{3}-\mathbb{J}_1\right) = \frac{\lambda_1^2+\lambda_2^2+\lambda_3^2-\lambda_1\lambda_2-\lambda_1\lambda_3-\lambda_2\lambda_3}{3}\left(\frac{\mathbb{J}_2}{3}-\mathbb{J}_1\right).
  \end{equation*}
\end{exam}

\bibliographystyle{abbrv}
\bibliography{biblio}
\end{document}